\documentstyle[11pt,psfig]{article}
\begin{document}
\title{\bf Statistical Investigations of 1315 Radio Pulsars}

\author{O. H. Guseinov$\sp{1,2}$
\thanks{e-mail:huseyin@pascal.sci.akdeniz.edu.tr}
, E. Yazgan$\sp2$
\thanks{e-mail:yazgan@astroa.physics.metu.edu.tr}
, S. \"{O}zkan$\sp1$
\thanks{e-mail:sozkan@pascal.sci.akdeniz.edu.tr} 
, A. Sezer$\sp1$
\thanks{e-mail:sezer@sci.akdeniz.edu.tr}
, and S. Tagieva$\sp3$
\thanks{e-mail:SO$\_$Tagieva@mail.ru} \\
\\{$\sp1$Department of Physics, Akdeniz University,} \\
{Antalya, Turkey} \\
{$\sp2$Department of Physics, Middle East Technical University,} \\ 
{Ankara, Turkey} \\
{$\sp3$Academy of
Science, Physics Institute Baku 370143,} \\ {Azerbaijan Republic}} 

\date{}
\maketitle

\begin{abstract}
\noindent
In this paper we have used the data of 1315 pulsars (Guseinov et al. 2002) 
for 
statistical analysis and showed the changes in pulsar parameters since   
the appearance of Taylor et al. (1996) catalog. Here we present the space 
distribution of pulsars, dispersion measures, 
distances from the galactic plane and from the Sun, 
electron density, luminosity distributions at 
400 and 1400 MHz, the relations between luminosities of pulsars at 400 
and 1400 MHz and the dependence of luminosity on age and on magnetic field. 
We also present the updated P-\.{P} diagram. \\ \\ KEY WORDS Pulsars, 
Compact Stars, Neutron Stars

\end{abstract}

\parindent=0.2in

\section{INTRODUCTION}

Observational and derived data of radio pulsars (PSRs) were last compiled 
in 1996 (Taylor et al. Catalog of 706 PSRs, unpublished 
work). This catalog has tremendous contributions in the development of PSR 
astronomy. 
The first full catalog (Manchester \& Taylor 1981) 
includes 333 PSRs which were discovered 
before 1981. Many PSRs were discovered after 1981 (Dewey et al. 
1985; 
Stokes et al. 1985, 1986; Clifton et al. 1992; Johnston et al. 1992a,b) 
and these PSRs were included in the Taylor et al. (1996) catalog.

After the publication of the
Taylor et al. (1996) catalog, there have been a 
number of pulsar surveys 
(Johnston et al. 1995; Lyne et al. 2000; D'Amico et al. 2001; Edwards \& 
Bailes 2001; Camilo et al. 2000; Manchester et al. 1996; Lyne et al.1998, 
Sandhu et al., 1997).
Furthermore, the inner parts of SNRs were 
scanned (completely or partially) to detect PSRs genetically connected with 
SNRs (Gorham et al. 1996; Kaspi et al. 1996; Lorimer et al. 1998). 
After 1996, it became more or less certain that the PSR-SNR pairs 
J2229+6114 -- G106.6+2.9  (Halpern et al. 2001), J0205+64 -- G130.7+3.1  
(Murray et al. 2002), J1846-0258 -- G29.7-0.3 
(Gotthelf et al. 2000), 
J1124-5916 -- G292.0+1.8 (Camilo et al. 2002), 
J1119-6127 -- 
G292.2-0.5  (Crawford et al. 2001; Pivovaroff 2001), J1803-2137 -- G8.7-0.1 
(Kassim \& Weiler 1990, Finley \& Oegelman 1994) 
and J1952+3525 -- G69.0+2.7 (Strom 1987, Shull et al. 1989)  
possibly have genetic 
connections (Allakhverdiev et al. 1997, Kaspi \& Helfand 2002). 
Also, there have been searches for PSRs in Globular Clusters 
(GC) (Camilo et al. 2000; D'Amico et al. 2001; Lyne 1995; Kulkarni \& 
Anderson 1996; Biggs and Lyne 1996).
Before 1996, 10 PSRs were detected in GC NGC104 (47 Tuc) and after 1996, 10 
new pulsars have been detected in this GC (Camilo et al. 2000). No new 
PSRs have been found from other GCs which have 
PSRs detected before 1996. After 1996, one PSR for each of the 
following GCs were detected; NGC 6266, NGC 6342, NGC 6397, NGC 6544 and NGC 
6752. However, the available
catalog (Taylor et al. 1996) has not been 
updated since then. 

Earlier, most of the PSRs were discovered at 400 MHz and 
frequencies close to 400 MHz. Since the dispersion measure (DM) of distant 
PSRs are larger, new PSR surveys and searches in SNRs and GCs have 
been conducted at 1400 MHz to obtain narrow pulses. Observations at 1400 MHz 
are also necessary for the study of other parameters of PSRs such as 
the spectral index.  
As expected, most of the new detections of PSRs 
were in the Galactic center direction. After 1996 no PSRs have been detected 
in the Magellanic Clouds (MC), except one PSR which has a period of 16 ms 
(PSR 
J0537-6910) detected in SNR N157B in the Large Magellanic Cloud (LMC), in 
the X-ray band (Marshall, Gotthelf, Zhang et al. 1998; Mignani, 
Pulone, Marconi et al. 2000). The reason for this PSR not to be seen in 
the radio band might be that it might 
have a medium radio luminosity. The number of PSRs detected in GCs 
increased by a factor of 1.5. The number of millisecond PSRs (P$<$0.1 sec 
and $\dot{P}<$10$^{-16}$ s/s) whose period and period derivative values 
are known, increased  
more than  1.5 times. The number of PSRs with low flux have considerably 
increased as a result of   
more (spatially) precise scans with high sensitivity. In the Arecibo window 
(40$^o<l<65^o$; $|$b$|<$2.5$^o$) (Hulse \& Taylor 1974, 1975), 12 new 
PSRs are discovered. 

Guseinov et al. (2002) 
compiled observational and derived data for 1315 PSRs.
The purpose of this work  is to  
perform statistical analysis using these data.   
Statistical investigations of PSRs have an important role in 
understanding their physical properties and origin. But today radio 
pulsars are not the unique phenomenon for neutron stars. 
Unexpected but very important discovery of the soft gamma repeaters and the
related objects, anomalous X-ray pulsars, showed us that we have problems 
in understanding neutron stars as a whole class. 
The different NS types might be due to the differences of their 
progenitor stars which we do not know very clearly. Statistical 
investigations of PSRs will help to understand the whole family of 
neutron stars. 

\section{SPACE DISTRIBUTION OF PULSARS}
The distribution of 1315 PSRs in the Galaxy with respect to 
Galactic longitude (l) and latitude (b) is displayed in Figure 1a.  
Comparing the new distribution with the space distribution of PSRs known 
before 1996 (G\"{o}k et al. 1996) we see that at high latitude 
($|$b$|>$10$^o$) many new 
PSRs have been detected both in the Northern and Southern sky. Naturally, 
many new 
PSRs are detected in the spiral arms and galactic center directions.  
If we compare the l-b distribution of 862 galactic 
PSRs in Figure 1b which have flux values measured at 1400 MHz, with the 
l-b distribution 
for PSRs with known flux at 400 MHz (G\"{o}k et al. 1996) then we see that  
for many PSRs with high galactic latitude, 1400 MHz flux values have been 
measured after 1996. But
comparison of Figure 1a and 1b shows that still for many PSRs (453 of 
them) especially at high latitudes, 
the 1400 MHz flux are not known yet. From these two figures, it 
is also seen that 1400 MHz flux values for some of the PSRs on the Galactic 
Plane are not known. The PSRs without 1400 MHz flux are the ones 
whose 400 MHz flux are lower. In Figure 1b PSR J1239+2453 
(l=252.4$^o$,b=86.5$^o$) has the highest galactic latitude b and 
PSR J0134-2937 (l=230.2, b=-80.2) has the lowest value of 
b. These PSRs respectively have distances of 0.56 kpc and 1.5 kpc and 
luminosities of 34.5 mJy kpc$^2$ and 20.2 mJy kpc$^2$ at 400 MHz, 3.13 mJy 
kpc$^2$ and 5.39 mJy kpc$^2$ at 1400 MHz (Guseinov et al. 2002).

Figure 1c displays the l-b distribution of 420 PSRs discovered after 1996 
with known 1400 MHz flux. As seen from the figure, after 1996, 
in the southern hemisphere of the Galaxy more PSRs have been discovered. 
Moreover, it is seen that, survey of the arms of the Galaxy in the 
galactic center direction has been more precise.
In this figure J1313+0931 (l=320.4$^o$, b=71.7$^o$) has the highest 
galactic latitude and J2346-0610 (l=83.9$^o$,b=-64.8$^o$) has the lowest 
latitude. The distances of these PSRs are respectively 0.6 and 1.9 kpc, 
the luminosities at 400 MHz are 1.26 and 
39.72 mJy kpc$^2$ and at 1400 MHz are respectively 0.057 and 7.23 mJy 
kpc$^2$.

Before, the most precisely scanned region of the Galaxy was the Arecibo 
window (Hulse \& Taylor 1974, 1975).  
Up to 
1996, in this region, 48 PSRs had been discovered. Now the number of PSRs in 
this region 
is 60. For these recently discovered 12 PSRs, the flux values at 1400 MHz 
are 
less than or equal to 0.5 mJy. One of these PSRs, namely PSR J 1907+0918 
(l=43.02$^o$, b=0.73$^o$) has a highly flat spectrum (F$_{400}$=0.4 mJy, 
F$_{1400}$=0.3 mJy). This PSR is located at a distance of 7.67 kpc and have 
luminosities of L$_{400}$=23.5 mJy kpc$^2$ and L$_{1400}$=17.7 mJy kpc$^2$. 
For the other 
11 PSRs flux values at 400 MHz do not exist in Guseinov et al. 
(2002). For the 48 PSRs discovered 
before, only 4 of them have flux values at 400 MHz lower than 1 mJy. These 
data show that, even though the survey is thorough, precise and 
scanning is conducted at a higher frequency, it is hard to detect new 
PSRs in this region. 

The whole Arecibo window had not been scanned with 
the same quality before. The best scanned region was 43$^o<$l$<$55$^o$ 
(Hulse \& Taylor 1974, 1975). Now the longitudes are scanned more precisely. 
Therefore it is natural that most of the recently discovered PSRs have 
galactic longitude 
values between 40$^o$ and 46$^o$ and most of them lie near to 40$^o$.

Figure 2a displays the DM values of 1312 PSRs in the Galaxy with respect to 
galactic longitude. None of the DM values of PSRs discovered up to 1996 
(except 
PSR J 1801-2306 (l=6.8$^o$, b=-0.8$^o$)which has DM=1074 cm$^{-3}$ 
pc) do not exceed 800 cm$^{-3}$ pc.  
Only 7 PSRs have DM$>$600 cm$^{-3}$ pc in Taylor et al. (1996) catalog. But 
now, as seen from Figure 2a, 83 PSRs have DM$>$600 cm$^{-3}$ pc and 31 
PSRs have DM$>$800 cm$^{-3}$ pc. 
These, naturally, are located at the arms close to the Galactic center. 
Four PSRs have DM$>$1100 and are located in 330$^o<$l$<$337$^o$. Among these 
four PSRs, the one with the
highest DM value (DM=1209 cm$^{-3}$ pc) is PSR J 1628-4848 
(l=335.5$^o$, b=0.2$^o$). The PSRs named in Figure 1: PSR 
J1801-2306 (l=6.8$^o$,b=-0.8$^o$), PSR J2044+46 (l=85.4$^o$, b=2.1$^o$), PSR 
J0631+1036 
(l=201.2$^o$, b=0.5$^o$), PSR J0855-4658 (l=267.3$^o$, b=-1$^o$) and PSR 
J1019-5749 
(l=283.8$^o$, b=-0.7$^o$) have the largest DM values in their own directions.

Figure 2b displays the distance from the galactic plane (Z) versus DM for 
1273 PSRs
with $|$Z$|<$2 kpc. The 83 PSRs having DM greater than 600 are all in 
Figure 2b.
The distances of these PSRs from the Galactic plane, 
do not
exceed 320 pc (Compare Figures 2a and 2b). The reason that the PSRs with 
high DM value have such low $|$Z$|$ values is
that the new surveys are focused on low galactic latitudes. However, 
even if 
high Galactic latitudes were scanned, we would not expect to find the ratio 
of PSRs
with DM values greater than 600 cm$^{-3}$ pc to all PSRs to change 
considerably. Because, as we move 
away from the galactic plane, electron density decreases abruptly. 
The PSRs, which are the members of the Galaxy but far away from the galactic 
plane, are located in GCs. Among these, the highest one from the galactic 
plane is PSR J1312+1810 (l=332.9, b=79.8) and have Z=18.6 kpc.

Figure 3a represents the distances of 1307 PSRs which belong to our Galaxy
versus galactic longitude (d-l). 
The farthest PSR from the Solar system is 
PSR J1312+1810. This PSR is a member of the globular cluster M53  
which have a very small value of 
DM=24 cm$^{-3}$ pc. 
The name of the most distant PSRs for several different longitude 
intervals are indicated in Figure 3a. These galaxy field PSRs (which do not 
belong to GCs) with distances d$>10$ kpc 
are all located close to the galactic plane (compare Figures 3a and 2b).
The PSRs with largest values of DM (Figure 2a) added here are mentioned 
before as the PSR in M53 and PSR J1305-6256 (l=304.5$^o$, b=-0.1$^o$).

Figure 3b represents the 
projection of the location of PSRs on 
the plane of the Galaxy. In this figure, the distance between consecutive 
rings is 2 kpc. From the figure it is seen that 
behind the Galactic center (l$\approx \pm 10^{o}$, d$>$ 8.5
kpc) there are not many PSRs with large distances similar to the PSRs, for 
example, in the direction l=300$^o$-310$^o$. The most important reason
is that near the Galactic center the background radiation is strong. 
Moreover as the distance of the PSR increases, DM 
increases, flux decreases and the $observed$ pulse width increases. 
Therefore it is hard to find such 
PSRs beyond the Galactic center.
For the distant PSRs in this direction, the DM values are not 
as high as the ones at l=340$^o$ (see Figure 2a) since background 
radiation do not let us see such distant PSRs even though their 
luminosity are large enough to be seen if they were not located near the 
center direction.

Only old population stars (stars with ages about 10$^{10}$ years) 
can come to dynamical stability. But old and young populations are not in 
dynamical equilibrium with each other, because the scale height and 
density distributions of the two populations are different. 
The total mass of gas and young stars located in the Galactic arms is 
only about 1$\%$ of the mass of the whole Galaxy and the parameters of these 
arms change with time. 
SFRs are yet more unstable having ages about an order of magnitude less 
than the characteristic ages of the arms of the Galaxy. Here, the arms 
are neither  in dynamical equilibrium with the Galaxy, nor with the SFRs. 
Therefore not in all 
regions on the geometrical plane, arms must be coincident with 
the geometrical 
plane of the Galaxy. For some regions SFRs may be higher or lower from 
the plane of 
the the Galaxy. From optical observations of cepheids with high 
luminosities (with long pulsation periods) and red supergiants, which are 
located at distances about 5-10 kpc from the Sun, it is found that, in the 
direction 
l=220$^o$-330$^o$, SFRs lie about 300 pc below the galactic plane.  
In the
same distance, in the direction l=70$^o$-100$^o$ it is found that SFRs 
lie about 400 pc above the galactic plane, 
and the young objects closer than 3-5 kpc in the 
direction 
l=270$^o$-320$^o$ are located about 150 pc down the geometrical plane of 
the Galaxy (Berdnikov 1987). 

Figure 4a displays the distribution of the distance of 1307 galactic PSRs 
from the plane of the Galaxy as a function of galactic longitude. Due to the 
fact that the 
PSRs in GC M5 (l=3.9$^o$, b=46.8,$^o$ d=9 kpc), M13 (l=59.8$^o$, b=40.9$^o$, 
d=7.7 kpc), 
M53 (l=332.9$^o$, b=79.8$^o$, d=18.9 kpc), M15 (l=65.1$^o$, b=-27.3$^o$, 
d=10 kpc) and 47 Tuc (l=305.9$^o$, b=-44.9$^o$, d=4.5 kpc)
have $|$Z$|>$3 kpc, they are plotted at closer distances than their real 
distance values. (There is
one PSR known in M53, 2 in M5, 2 in M13, 8 in M15 and 20 in 47 Tuc). In 
Figure 
4a, the values of distances of these PSRs from the plane of the Galaxy  
are shown. Naturally, the PSRs with low flux are not scanned thoroughly 
with the same precision on all of the sky. 
This and the deviation of the SFRs from the galactic plane
affect 
mostly the distant PSRs which are located   
in the peripheric parts of the Galaxy, so we have 
plotted the Z-l distribution of  551 
PSRs with d$>$5 kpc in 
Figure 4b. As it is seen from the figure, expectedly, the PSRs in the 
direction l=70$^o$-100$^o$ are generally
located above the galactic plane and the PSRs in the 
direction l=200$^o$-300$^o$ are located below the Galactic plane. 
Among these PSRs J1721-1936 (l=4.9$^o$, b=9.7$^o$), J0952-3839 
(l=268.7$^o$, b=12.3$^o$) and J0745-5351 (l=266.6$^o$, b=-14.3$^o$) have 
the largest deviations from the plane of the Galaxy and they have distances 
of 9, 7.2 and 6.2 kpc, respectively.

Figure 5 represents the average value of electron density, n$_e$, 
along the line
of sight as a function of galactic longitude (1312 PSRs; 5 PSRs in 
MCs are included).  
PSR J1818-1519 (l=15.6$^o$, b=0.1$^o$, n$_e$=0.085 cm$^{-3}$) has a 
distance of 8.2 kpc, J0826+2637 (l=196.9$^o$, b=31.7$^o$, n$_e$=0.049 
cm$^{-3}$) has 
d=0.4 kpc, PSR J0828-3417 (l=253.9$^o$, b=2.6$^o$, n$_e$=0.039 cm$^{-3}$ has 
d=1.2 
kpc, PSR J1119-6127 (l=292.2$^o$, b=-0.54$^o$, n$_e$=0.094 cm$^{-3}$) has 
d=7.5 kpc, 
PSR J1302-6350 (l=304.2$^o$, b=-0.9$^o$, n$_e$=0.113 cm$^{-3}$) has d=1.3 
kpc, PSR J1644-4559 (l=339.2$^o$, b=-0.2$^o$, n$_e$=0.107 cm$^{-3}$) has 
d=4.5 kpc. 
PSR J0835-4510 (l=263.6$^o$, b=-2.8$^o$) in Vela SNR has the 
highest value of n$_e$=0.15 cm$^{-3}$. 
This value is found by assuming the distance to the PSR to be d=0.45 kpc. 
In the following paragraph we explain our reasons to take the distance to 
Vela to be 0.45 kpc.

The distance to the Vela remnant is given as d=250 pc (\"{O}gelman et 
al. 1989), d=250$\pm$30 pc 
(Cha et al. 1999; Danks 2000), d$\approx$280 pc (Bocchino et al. 1999). 
But we have to take into account the fact 
that Vela SNR is
expanding in a dense medium. The magnetic field is given as 6$\times 
10^{-5}$ 
Gauss (de Jager et al. 1996) and as (5-8.5)$\times 10^{-5}$ Gauss 
(Bocchino et al. 2000) and explosion energy is (1-2)$\times 10^{51}$ 
ergs (Danks 2000). These values are more than the average. 
Naturally, the errors are high for these quantities. 
However, if we assume that these data are correct and 
the distance to the Vela remnant is 250 pc, then 
in $\sum$-D diagram Vela and the SNR G 327.6+14.6 are located at the same 
position. However, this is not convincing, since Vela and SNR G 
327.6+14.6 are formed from different type of SN explosions, i.e. SNR G 
327.6+14.6 was formed from a type Ia explosion (Hamilton et al. 1997). 
This SNR has a distance of 500 pc from the galactic plane, and 
consequently it lies in a much less dense medium than the medium of Vela 
remnant. And in the $\sum$-D diagram it is hard to understand the deviation 
of the Vela remnant from the $\sum$-D relation. On the other hand, the 
deviation of SNR G
327.6+14.6 is readily understandable, because this SNR is formed from a 
type Ia explosion, but the $\sum$-D relation is for S and C-type SNRs.
The distance of OB associations including the young open clusters (OC) in 
the direction of Vela remnant are not closer than 0.25 kpc.   
One of these OCs which lie in the OB association (OB2) closer to the Sun, in 
the direction of Vela, is Pismis 4 (l=262.7$^o$, b=-2.4$^o$). Pismis 4 
has a well known distance of 0.6 kpc.  
(Ahumada \& Lapasset 1995; Ayd$\i$n et al. 1997). Since the 
progenitors of 
SNRs (or pulsars) are massive stars, the probability that Vela is not at 
0.25 kpc but closer to the SFR is high.
As seen from Figure 5, PSR J1302-6350 has the second highest n$_e$ (after 
Vela) which is 0.113 cm$^{-3}$ is 
PSR J1302-6350 (l=304.2$^o$, b=-0.9$^o$) which has a Be star 
companion (having a variable wind) and which is at a distance of 1.3 
kpc (Johnston et al. 1994). So, it is normal for this PSR to have such a 
high n$_e$.
The third highest n$_e$ is 0.107 cm$^{-3}$ which is calculated for PSR J 
1644-4559 (l=339.2$^o$,b=-0.2$^o$). Since the luminosity at 1400 MHz for 
this last PSR is 
higher than all other PSRs with known flux (L$_{1400}$=6.3$\times 10^3$, 
L$_{400}$=7.9$\times 10^3$, see Figure 6b), it 
would not be true to accept its distance to be more than 4.5 kpc. 
PSR J1341-6220 (l=308.7$^o$, b=-0.4$^o$) has 
n$_e$=0.091 (which is the forth highest n$_e$) at d=8 kpc. 
For SNR G308.8-0.1 
which is thought to be genetically connected with this PSR, the 
angular size and flux value are not precisely 
known. But we cannot accept that this SNR is located farther. Due to 
these reasons the distance to the PSR-SNR pair must be about 8 kpc 
(Using the $\Sigma$-D diagram of Guseinov et al., in preparation). In the 
direction and distance of this PSR, SFR lies below the 
galactic plane. Consequently, n$_e$ shows asymmetric distribution about 
the plane, and the value of n$_e$ below is higher than the corresponding 
n$_e$ at the same longitude with same b value but above the plane. 
For all of the  
PSR sample, average value of n$_e$ is around 
0.04 cm$^{-2}$. Using 250 pc for the distance of Vela leads 
to an electron density two times higher than the electron density calculated 
assuming the distance to be about 400 pc. However, even when the distance 
is assumed to be 400 pc, n$_e$ is readily the highest of all.
So distance to Vela is at least 0.4 kpc and the most recent SNR 
catalog (Green, 2001) gives the distance to Vela as $\sim$0.5 kpc.

The value of the electron density along the line of sight is 
high for the Vela PSR because it lies 
in an SNR in the nearest OB association. There are about 15 PSRs 
genetically connected to SNRs in our Galaxy (Kaspi \& Helfand 2002). Some of 
these SNRs with high probability lie in OB 
associations. 
However, since these associations are distant sources (so column depth is 
high), value of n$_e$ along the line of sight is less.

Figure 6a displays the luminosity at 400 MHz (L$_{400}$=F$_{400}$d$^2$ 
mJy kpc$^2$) with respect to longitude 
for 685 PSRs. For 12 PSRs Log L$_{400}$$<$0 and 10 of these 
PSRs lie in 0$^o$$<$l$<$150$^o$.
 This is because the Arecibo survey scanned the 
northern sky more thoroughly with very high sensitivity. The PSRs with least 
L$_{400}$ 
are J0108-1431 (l=140.9,b=-76.8) and J0006+1834 (l=108.2,b=-42.9). They 
respectively have distances of 0.08 kpc and 0.7 kpc and luminosity  
values of 
0.058 mJy kpc$^2$ and 0.098 mJy kpc$^2$. The strongest PSRs at 400 MHz 
are respectively J0529-6656 (l=272.2, 
b=-32.8) belonging to the LMC, J0738-4042 (l=254.2, b=-9.2) and J1901+0331 
(l=37.2, b=-0.6). These have distances of 50, 8 and 8.03 kpc respectively. 
Luminosity values of these PSRs at 400 MHz are 1.37$\times 10^4$ mJy 
kpc$^2$, 
1.28$\times 10^4$ mJy kpc$^2$ and 1.06$\times 10^4$ mJy kpc$^2$, 
respectively.

In Figure 6b the luminosity at 1400 MHz versus longitude for 862 galactic 
PSRs are displayed. At these frequencies, the galactic plane is 
scanned 
with high sensitivity, and also the flux of closer PSRs which have known 
values at
 400 MHz are measured. From the
fact that on the average the flux of PSRs at 400 MHz is about 6-7 times more 
than
the flux at 1400 MHz, it could be inferred that the PSRs with Log L$_{400}$ 
$<$ 1 will
have values of Log L$_{1400}$ $<$ 0. For some of these PSRs (with 
Log L$_{400}$ $<$ 1) values of F$_{1400}$ are unknown. The number of
 PSRs with Log L$_{1400}$ $<$ 0 is 42 and naturally most of them have high
galactic latitudes since background radiation is low at high galactic 
latitudes (for approximately 30 of them, $|$b$|>$5$^o$ ). PSRs 
J0030+0451 (l=113.1$^o$, b=-57.6$^o$) and J2144-3939 (l=2.7$^o$, 
b=-49.5$^o$) have the lowest values of 
luminosity at 1400 MHz. They respectively are located at distances of 0.235 
and 0.2 kpc and have L$_{1400}$=0.033 and 0.032 mJy kpc$^2$. The strongest 
PSRs 
at 1400 MHz are J1644-4559 (l=339.2$^o$, b=-0.2$^o$), J0738-4042 
(l=254.19$^o$, b=-9.19$^o$), J1935+1616 
(l=52.4$^o$, b=-2.9$^o$) and J1243-6423 (l=302.5$^o$, b=-1.5$^o$) 
PSR J0738-4042 has a large luminosity also at 400 MHz. But these PSRs 
are 
not related to the PSRs which we have mentioned as the farthest PSRs in 
Figure 3a, i.e. 
the farthest PSRs do not necessarily have the highest luminosities;
we can observe the farthest ones because they have high galactic 
latitudes, and the OCs are scanned thoroughly. These PSRs respectively have 
distances of 4.5, 6.2, 8.3 and 9 kpc and  
luminosities at 1400 MHz 6.29$\times 10^3$, 5.5$\times 10^3$, 1.61$\times 
10^3$ and 1.63$\times 10^3$ mJy kpc$^2$.
 
\section{Luminosities of PSRs at 400 and 1400 MHz}

Figure 6c displays Log L$_{1400}$ versus Log L$_{400}$ values. 
The number of PSRs
with known values of flux both at 400 and 1400 MHz simultaneously  
is 449. The relation between L$_{1400}$ and L$_{400}$ we infer from  
these PSRs in Figure 6c is
\begin{equation}
        L_{1400}=0.1L_{400}^{1.0 \pm 0.1}
\end{equation}
PSRs have radio spectral index values, on the average, that the 
flux
at 1400 MHz is about 6.3 times less than the flux at 400 MHz. As seen from 
Figure 6c,
some PSRs deviate strongly from the relation in Equation 1. For example,
J1302-6350 (l=304.2, b=-0.9) which has a Be star companion has 
L$_{1400}$/L$_{400}$=1.35.
For PSR 
J1114-6100 (l=291.4$^o$, b=-0.3$^o$) value of L$_{1400}$/L$_{400}$=1.16,
 for PSR J1633-4805 (l=336.3$^o$, b=-0.11$^o$) 
value of L$_{1400}$/L$_{400}$=0.005. 
For the PSR J1644-4559 (l=339.2$^o$, b=-0.2$^o$)
which has the highest luminosity among all PSRs at 1400 MHz, value 
of L$_{1400}$/L$_{400}$=0.83.
Characteristic ages of these PSRs are respectively equal to 
3.3$\times 10^5$ yr, 2.9$\times 10^7$ yr, 3$\times 10^5$ yr, 1.4$\times 
10^5$, and 3.5$\times 10^5$ yr.

In Figure 7a luminosity values of PSRs at 1400 MHz versus their 
characteristic ages (Log L$_{1400}$-Log $\tau$ diagram) for 853 PSRs
(5 PSRs in MC are included) is displayed. From the figure it is seen that 
for PSRs whose
Log $\tau$ $>$6.5, luminosity decreases as $\tau$ increases.  
Here, the
effect of millisecond PSRs is strong. The luminosity of young PSRs, 
practically, do not depend on $\tau$.

In Figure 7b, luminosity at 1400 
MHz versus magnetic field  
(Log L$_{1400}$-Log B diagram) for the same 853 PSRs is given. Magnetic 
field is calculated from the well-known relation B=3.3$\times 
10^{19}(P\dot{P})^{1/2}$. As seen
from the figure the luminosity of PSRs, including millisecond PSRs with 
magnetic fields less than 3$\times10^{10}$ Gauss have considerably low 
luminosity values, on the average, than luminosities of young PSRs (i.e. B$> 
3\times 10^{11}$ G). As seen 
from the figure the PSRs with magnetic fields about 10$^{12}$ Gauss can have 
luminosity values in
a 5 order interval. Such PSRs with high magnetic fields can also have low
luminosities like millisecond PSRs have. As inferred from Figure 7a and b 
the huge  
difference in the luminosities of PSRs are innate. Most of the 
PSRs are born single. In Figure 8a Log L$_{1400}$-Log L$_{400}$ 
diagram for single PSRs with $\tau$ $<$ 10$^7$ years is represented. For the 
273 PSRs in 
this figure, the relation between L$_{1400}$ and L$_{400}$ is found as
\begin{equation}
        L_{1400}=0.17L_{400}^{0.99 \pm 0.1}
\end{equation}

From figures 6c and 8a we see that spread in the luminosities of young PSRs 
may be
 large (see the errors 
in slopes of expressions (1) and (2)). From Figures 6c and 8a we also see 
that PSRs with luminosities L$_{400}<$30 
mJy kpc$^2$ are in average considerably softer (L$_{1400}$/L$_{400}$ is 
small) than PSRs with higher luminosity values. 
Since PSRs are non-stationary objects, their flux and spectral index can
change in time. 
To have confidence, we have plotted Log 
L$_{1400}$-Log L$_{400}$ diagram for PSRs with $\tau < 10^6$ years 
(Figure 8b). The degree of dependence inferred from this figure:
\begin{equation}
L_{1400}=1.15L_{400}^{0.72\pm 0.05}
\end{equation}
for PSRs with $\tau < 10^6$ years is considerably less.

Radio luminosity and spectral properties of neutron 
stars are very important also for investigation of AXPs, SGRs and 
DRQNSs. So we have also investigated L$_{1400}$-L$_{400}$ dependence on 
magnetic field strength. However, we found that the slope of 
L$_{1400}$-L$_{400}$ do not depend on magnetic field strength. 
Evidently, there can be huge uncertainties in the flux
measurements (especially when the flux is low). Because of
these reasons, even though we do
not consider the distance errors, the location of several PSRs can 
change
in Figures 6c, 8a and 8b. However, we do not expect the Equations 1, 2 and 3
to change, since the number of data is high enough.

The P-\.{P} diagram is displayed in Figure 9. 
In this figure 
constant characteristic age line, constant magnetic dipole field lines and 
the constant rate of 
rotational energy loss lines are also plotted. The rate of rotational 
energy loss \.{E} is calculated from the formula \.{E}=3.94 
10$^{46}$\.{P}/P$^3$. As expected, the form of the diagram changed as the 
number of data increased.

\section{Discussion and Conclusions}
As the number of PSRs increases, we understand the physical properties 
of neutron stars more clearly, 
parameters of PSRs are better determined and the distances become more 
reliable. Therefore, our knowledge about PSRs considerably increased from 
their space-kinematic characteristics to their physical properties. It is 
especially necessary to note that the increase of our knowledge is due to 
the precise observations at 1400 MHz.

$Acknowledgements$.
We thank T\"{U}B\.{I}TAK, the Scientific and Technical Research Council of
Turkey, for support through TBAG-\c{C}G4.

\begin{figure*}
\centerline{\psfig{file=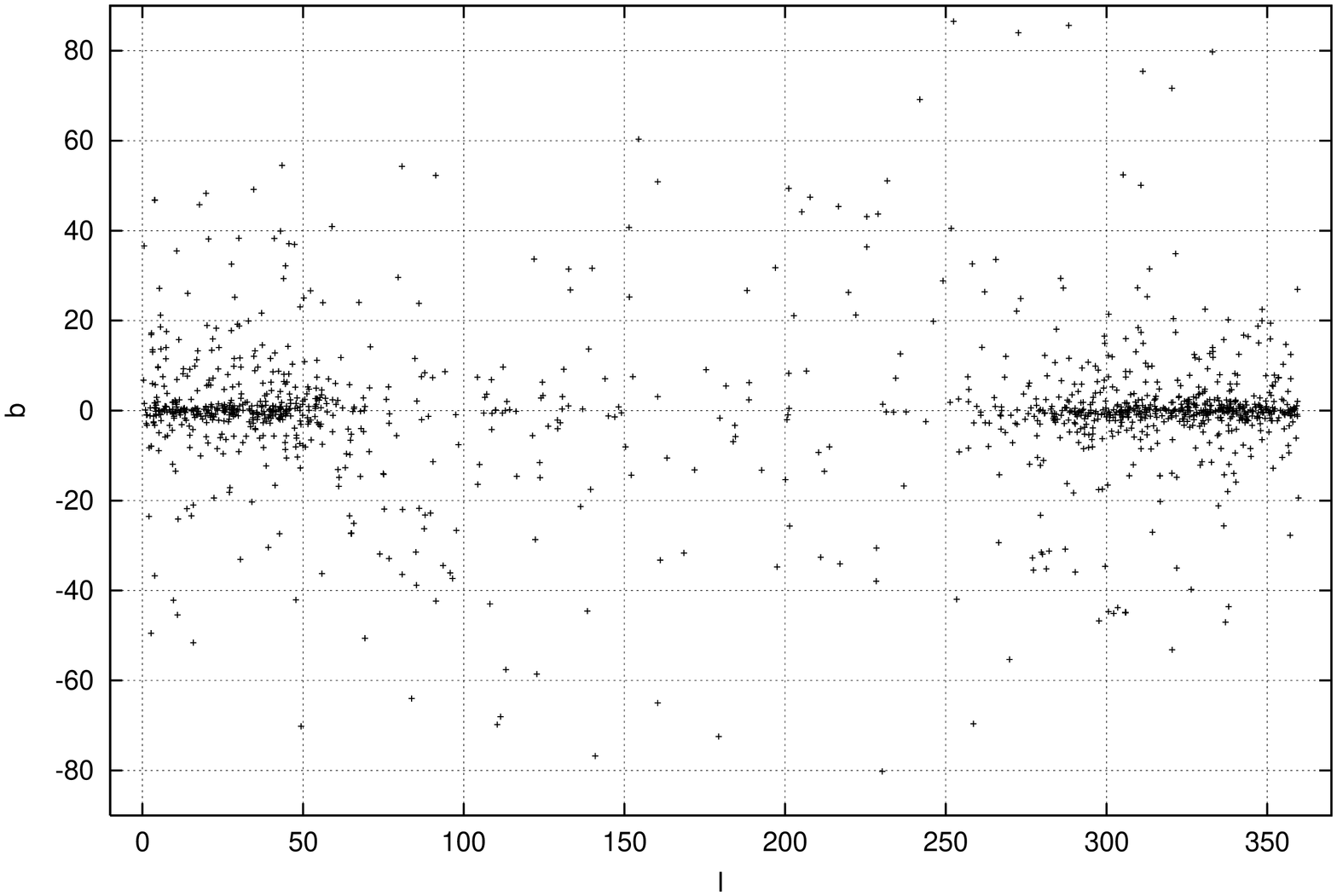,width=16cm,angle=-90}}
{Figure 1a. Galactic longitude (l) versus Galactic latitude (b) of 1315 
PSRs.} 
\end{figure*}

\begin{figure*}
\centerline{\psfig{file=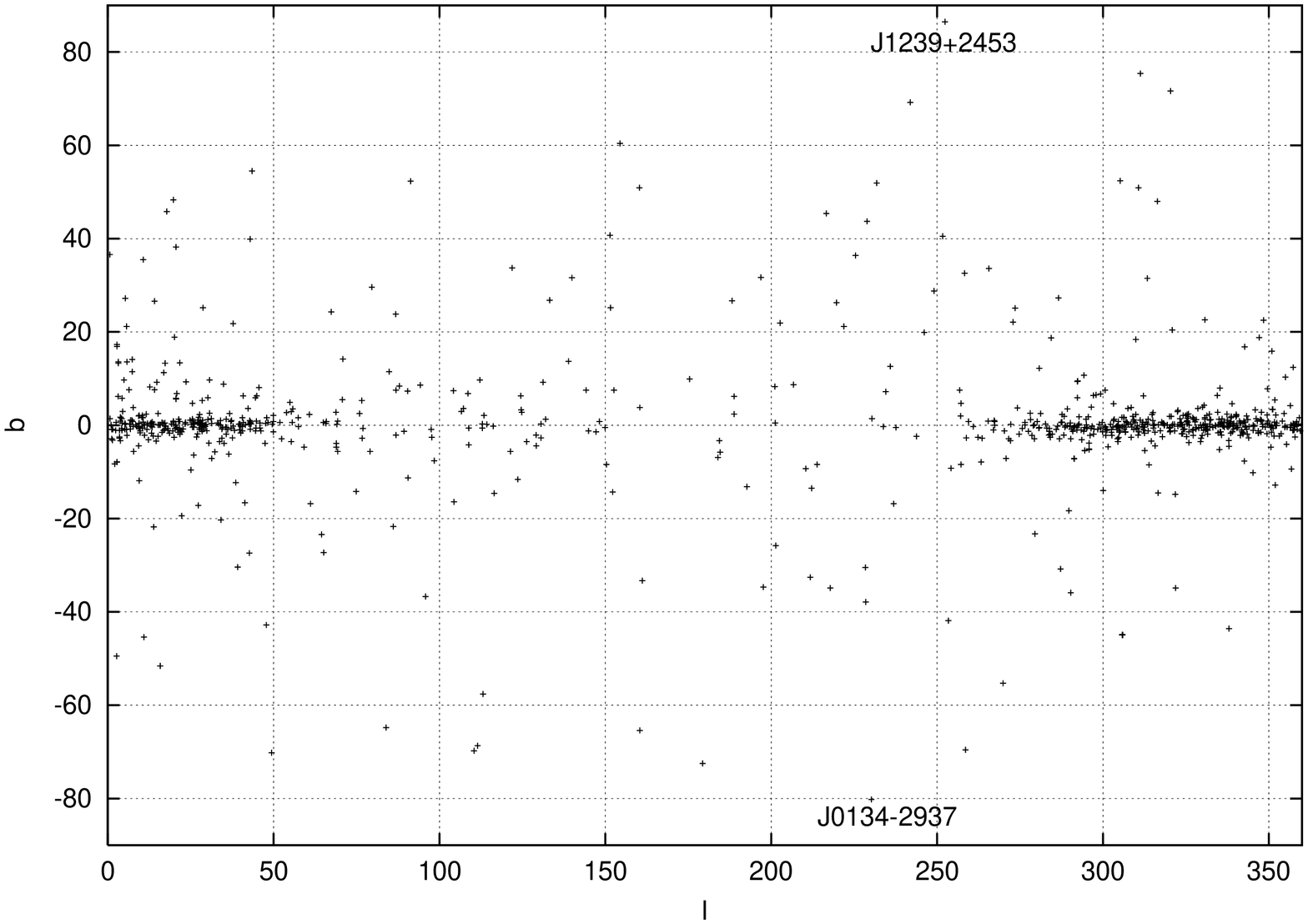,width=16cm,angle=-90}}
{Figure 1b. The l-b distribution of 862 galactic
PSRs which have flux measured at 1400 MHz.}                          
\end{figure*}

\begin{figure*}
\centerline{\psfig{file=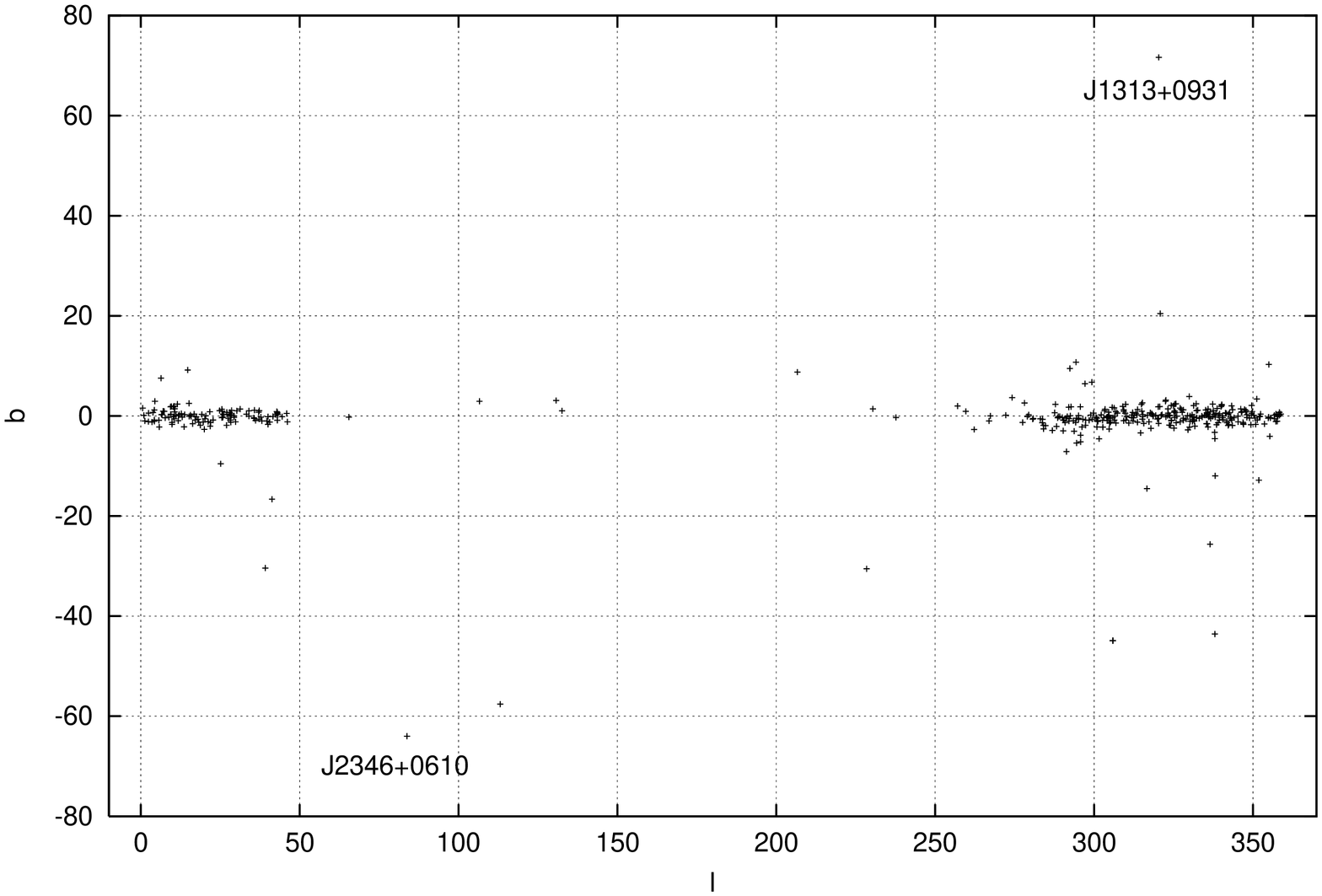,width=16cm,angle=-90}}
{Figure 1c. The l-b distribution of 419 PSRs 
discovered after 1996 with flux measured at 1400 MHz}   
\end{figure*}

\clearpage
\begin{figure*}
\centerline{\psfig{file=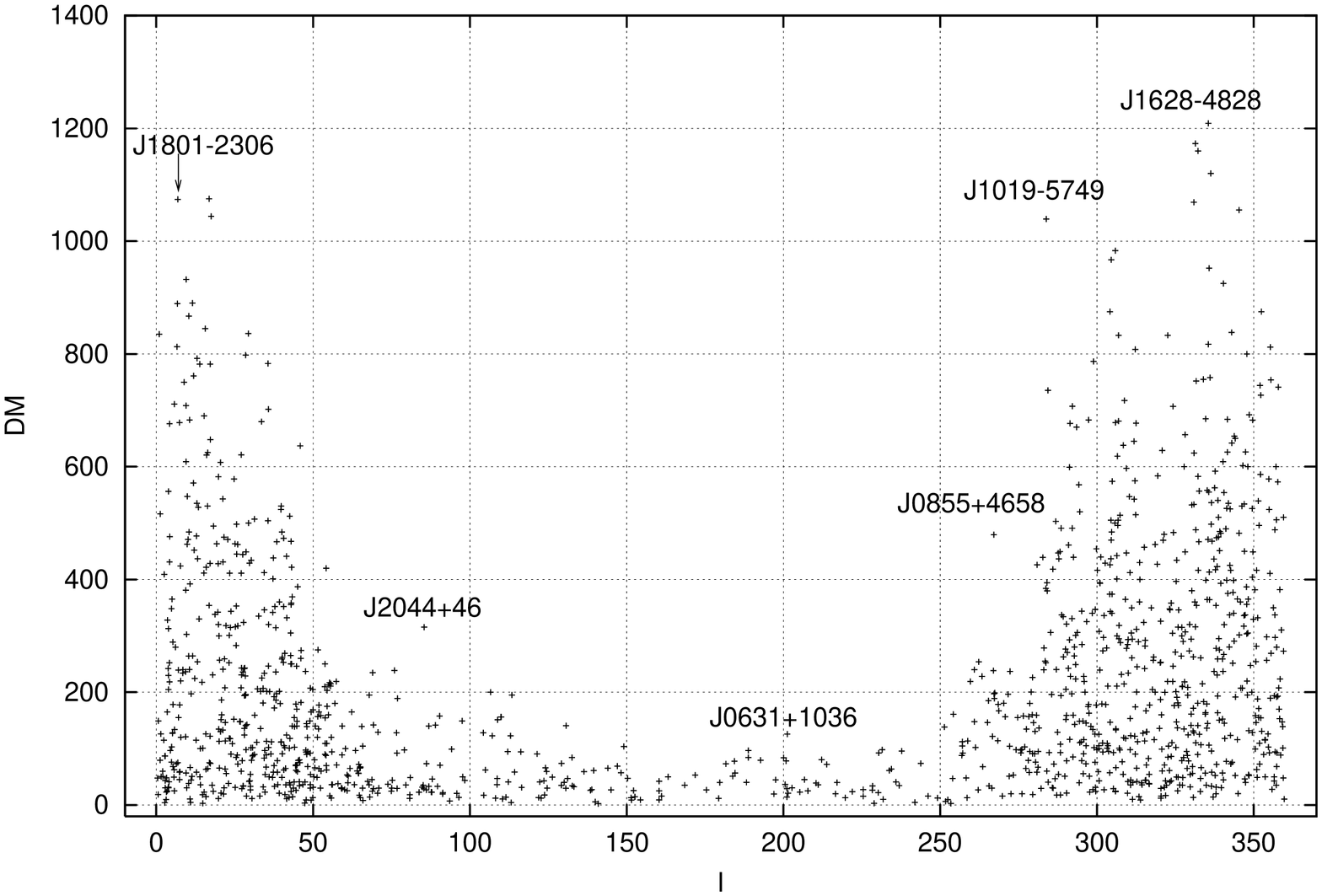,width=16cm,angle=-90}}
{Figure 2a. DM values of 1312 PSRs in the Galaxy with respect to
Galactic longitude.}
\end{figure*}

\begin{figure*}
\centerline{\psfig{file=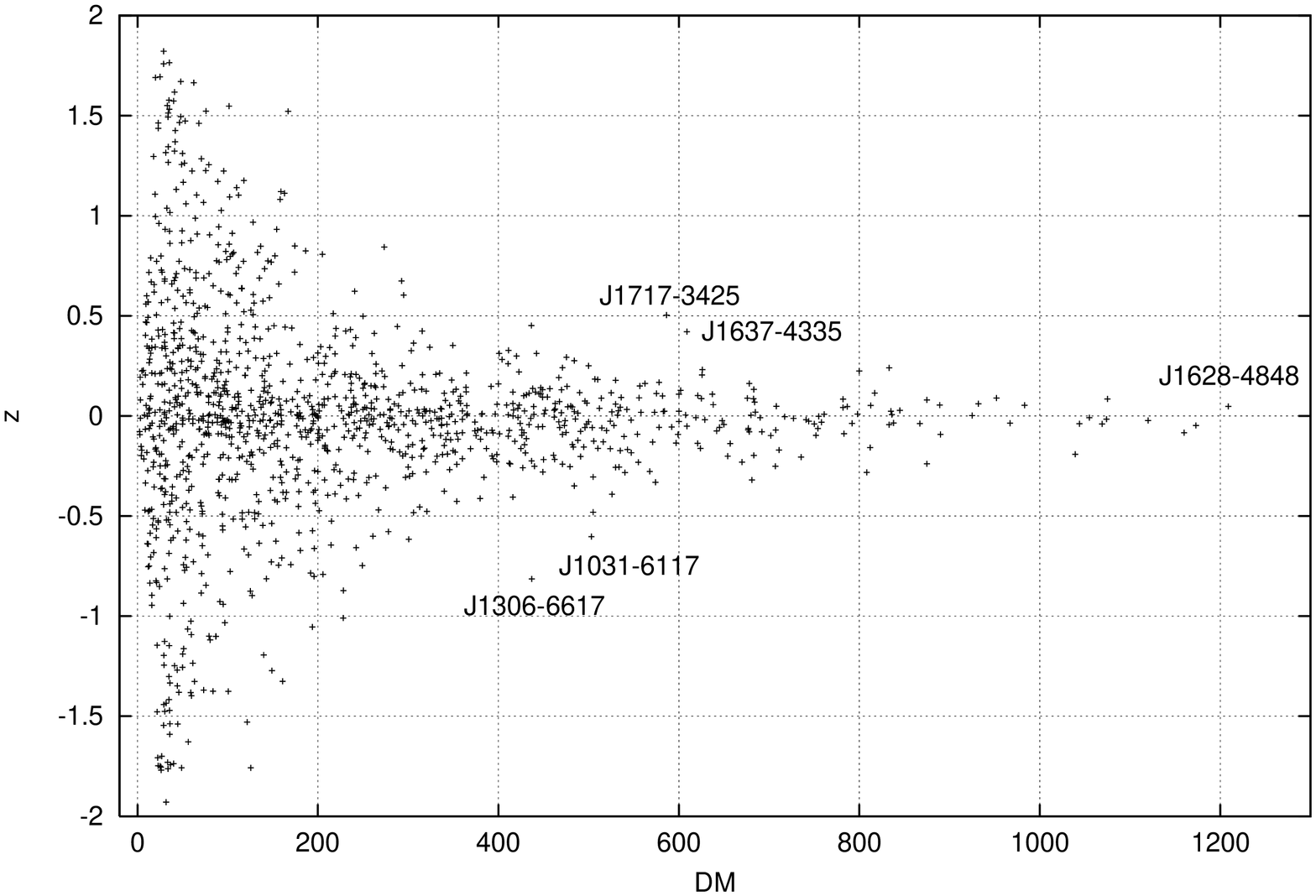,width=16cm,angle=-90}}
{Figure 2b. The distance from the Galactic plane vs. DM for 1273
PSRs with $|$Z$|<$2 kpc.}
\end{figure*}

\newpage
\clearpage

\begin{figure*}
\centerline{\psfig{file=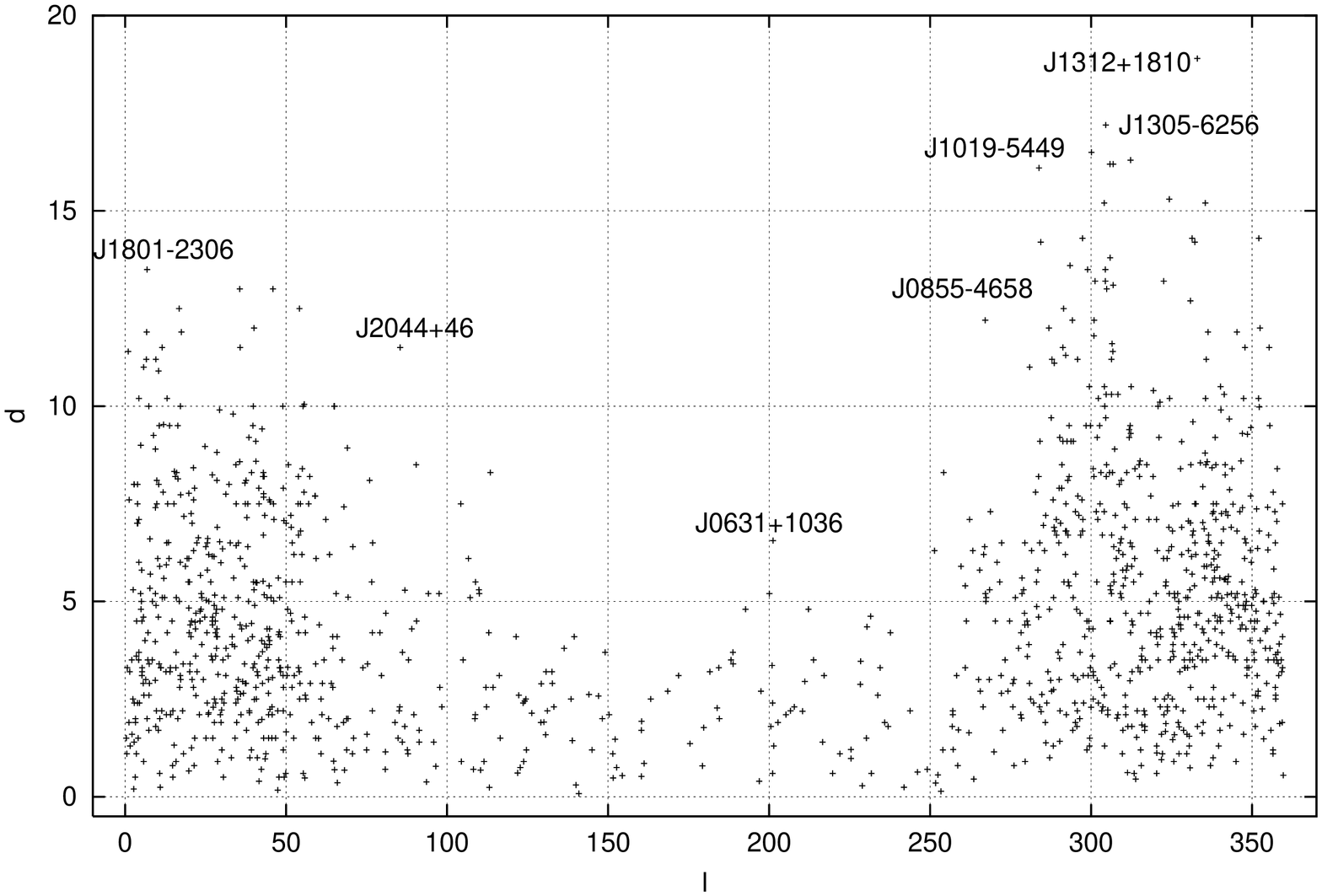,width=16cm,angle=-90}}
{Figure 3a. The distances of 1307 PSRs which belong to our Galaxy
versus Galactic longitude (d-l).}
\end{figure*}

\begin{figure*}
\centerline{\psfig{file=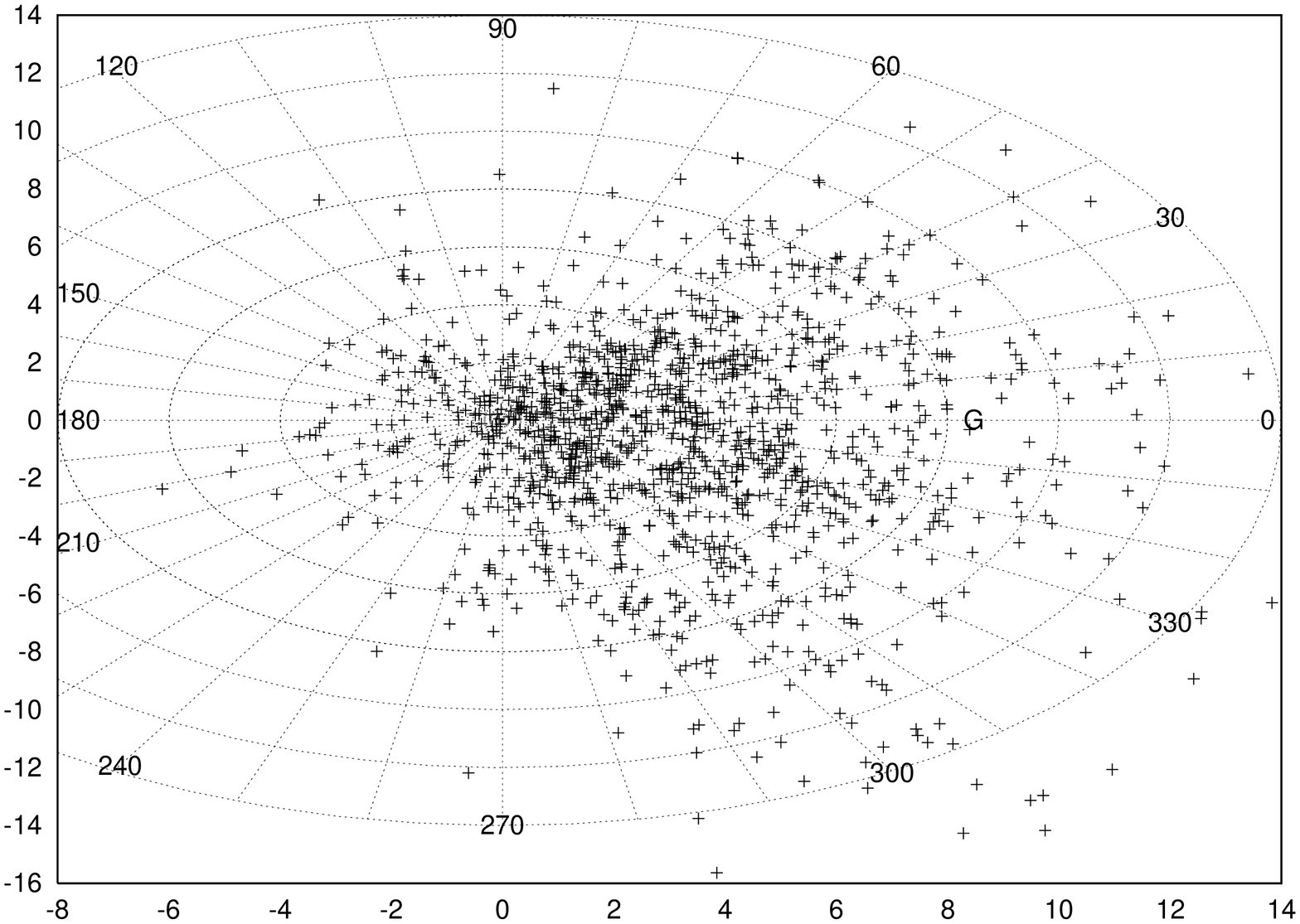,width=16cm,angle=-90}}
{Figure 3b. The projection of the locations of PSRs on      
the plane of the Galaxy. G represents the galactic center.}
\end{figure*}

\newpage
\clearpage

\begin{figure*}
\centerline{\psfig{file=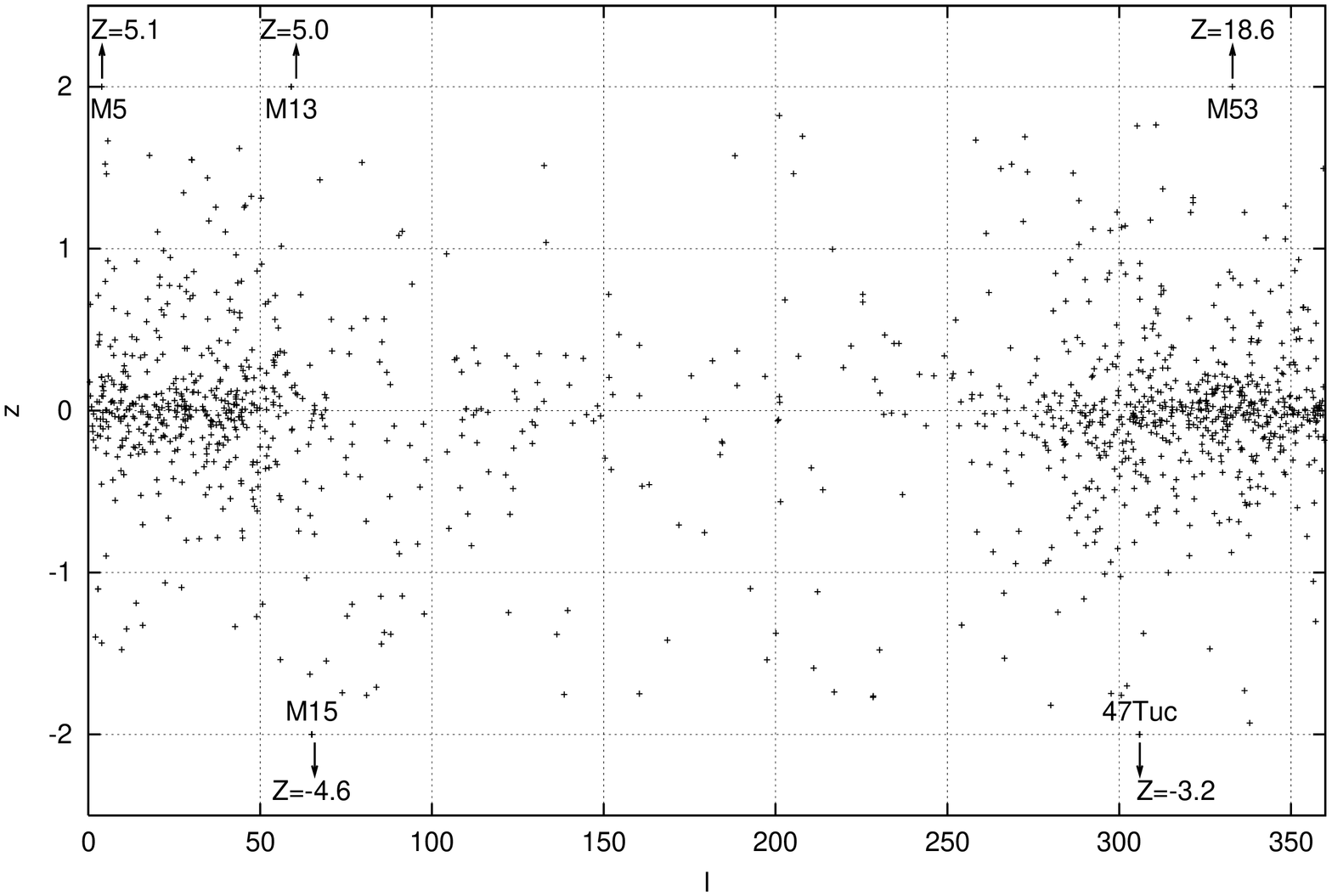,width=16cm,angle=-90}}
{Figure 4a. The distribution of the distances of 1307 galactic PSRs
from the plane of Galaxy as a function of galactic longitude.} 
\end{figure*}

\begin{figure*}
\centerline{\psfig{file=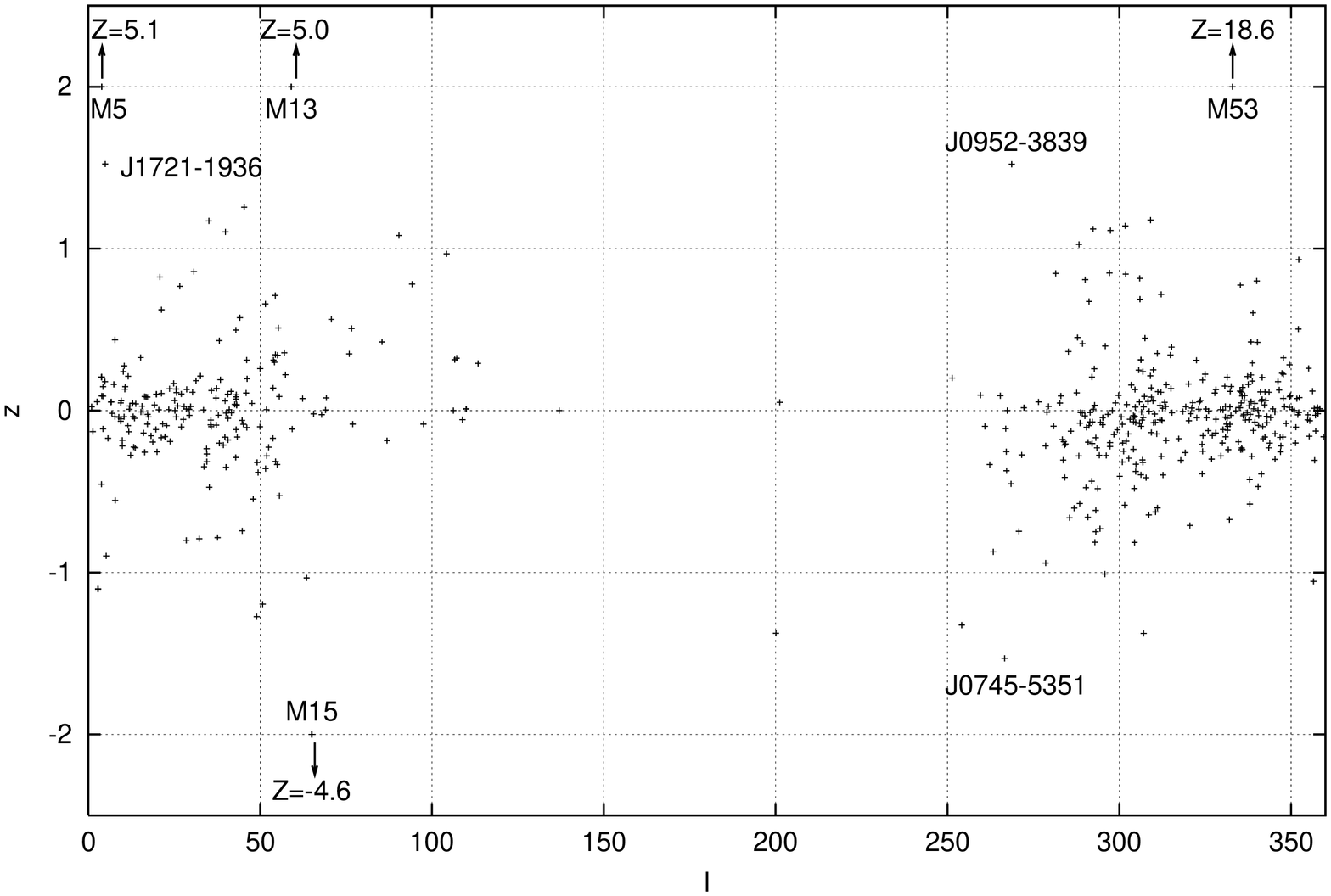,width=16cm,angle=-90}}
{Figure 4b. Z-l distribution for 551 PSRs with distance more than 5 kpc.} 
\end{figure*}

\newpage
\clearpage

\begin{figure*}
\centerline{\psfig{file=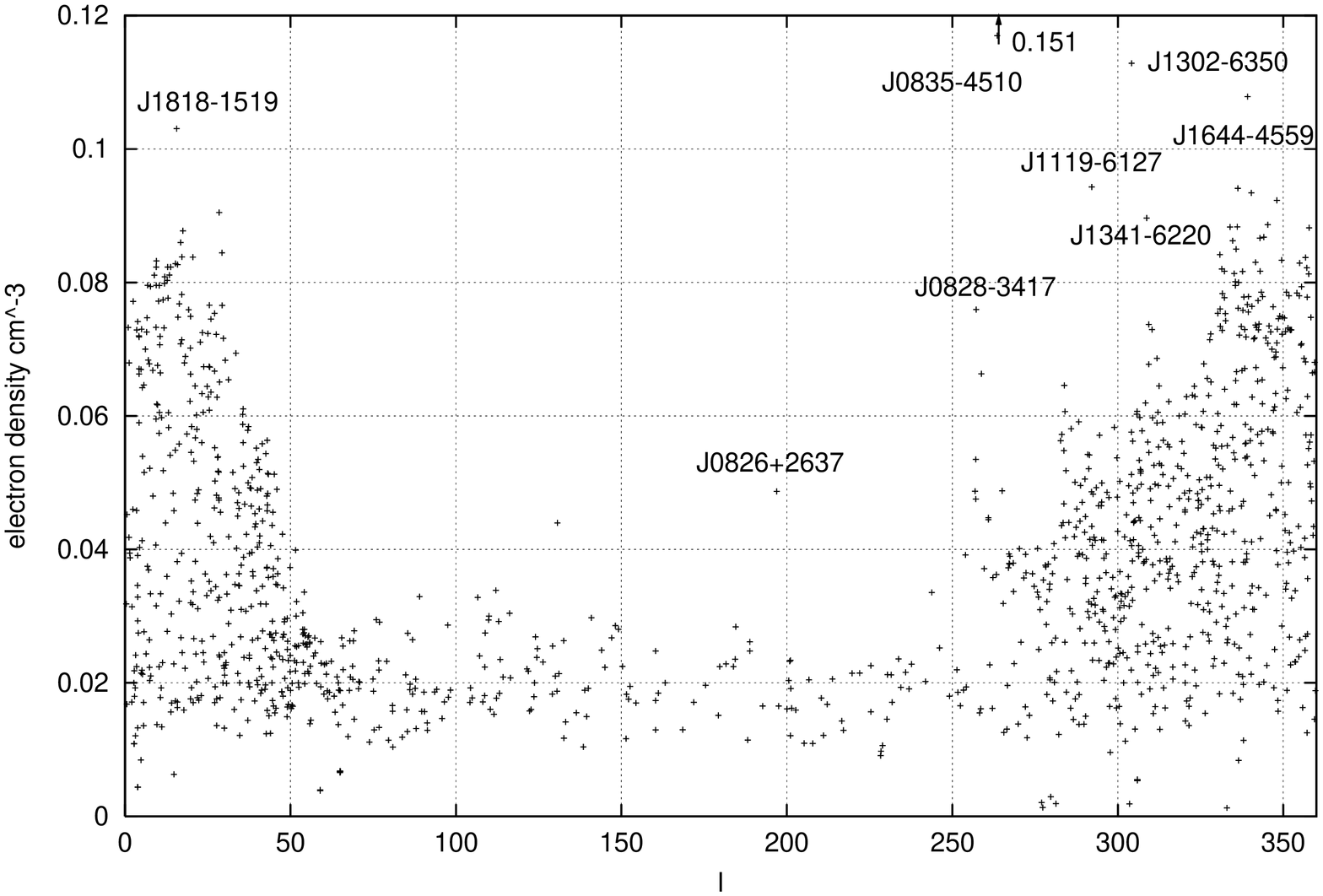,width=16cm,angle=-90}}
{Figure 5. The average value of electron density, n$_e$,
along the line
of sight vs. Galactic longitude for 1312 PSRs}
\end{figure*}

\newpage
\clearpage

\begin{figure*}
\centerline{\psfig{file=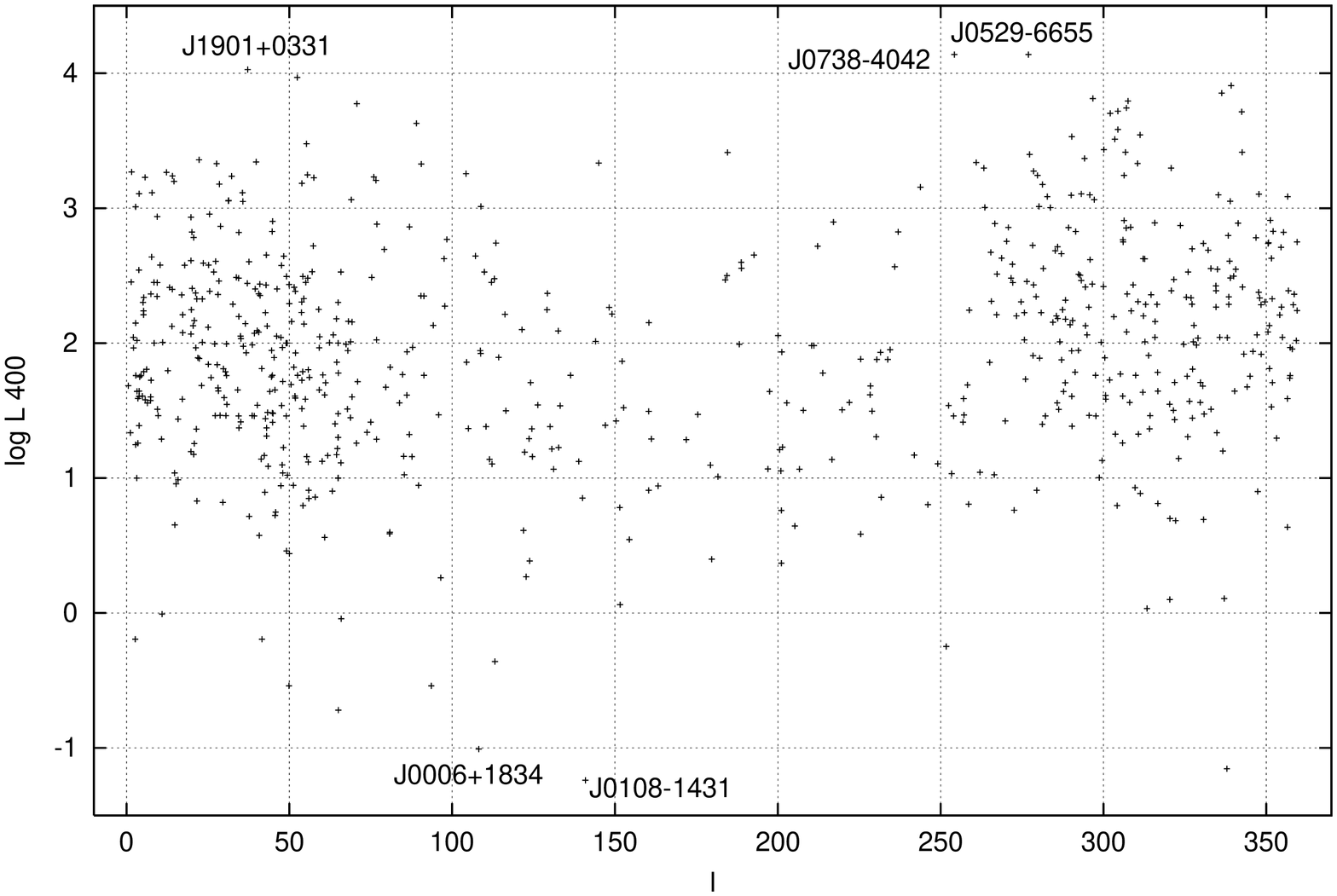,width=16cm,angle=-90}}
{Figure 6a. The luminosity at 400 MHz (L$_{400}$=F$_{400}$d$^2$
mJy kpc$^2$) vs. longitude
of 685 Galactic PSRs.}
\end{figure*}

\begin{figure*}
\centerline{\psfig{file=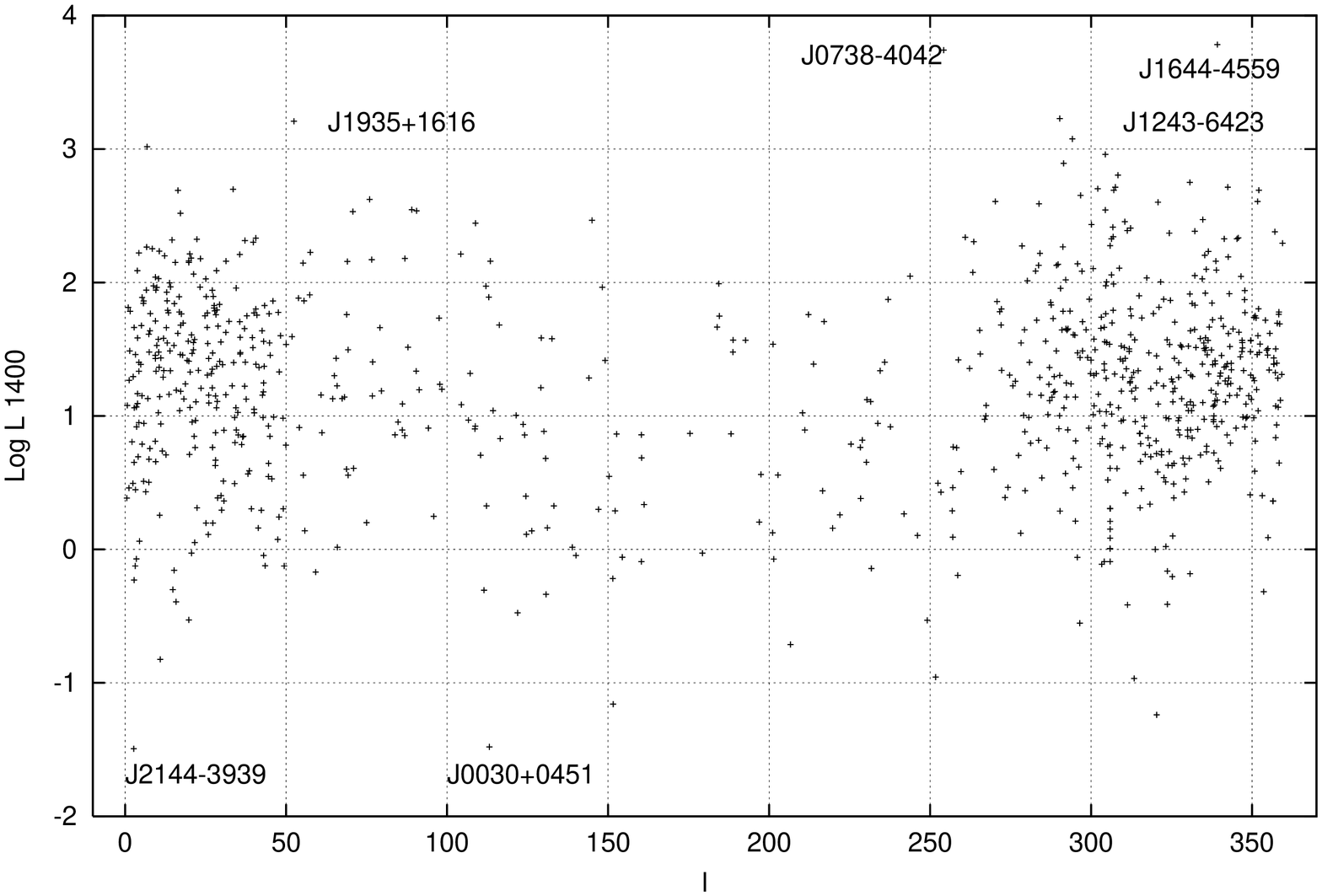,width=16cm,angle=-90}}
{Figure 6b. The luminosity at 1400 MHz vs. longitude of 862 Galactic PSRs}
\end{figure*}

\begin{figure*}
\centerline{\psfig{file=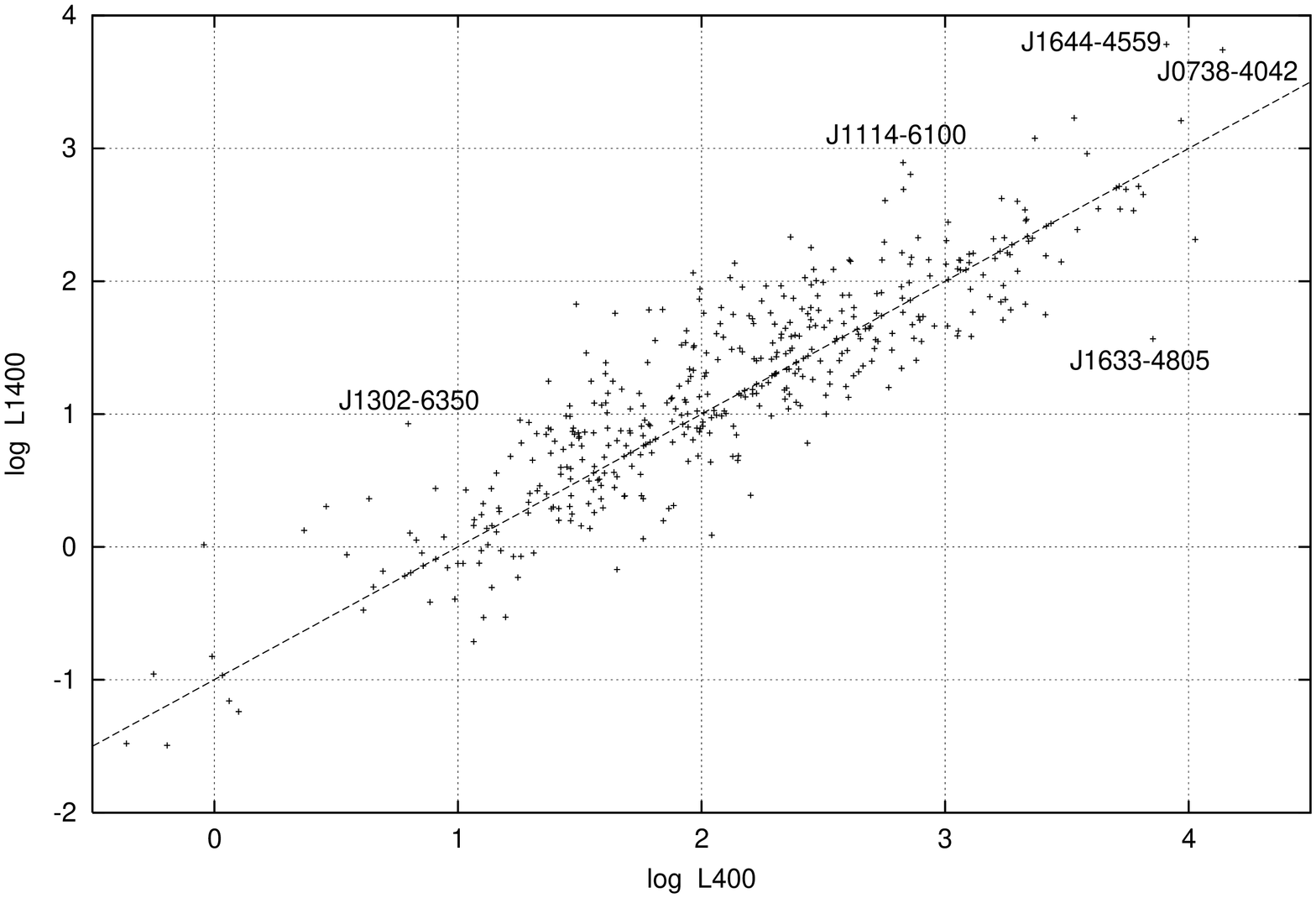,width=16cm,angle=-90}}
{Figure 6c. The values of Log L$_{1400}$ vs. values Log L$_{400}$  
for 449 PSRs}
\end{figure*}

\newpage
\clearpage
\begin{figure*}
\centerline{\psfig{file=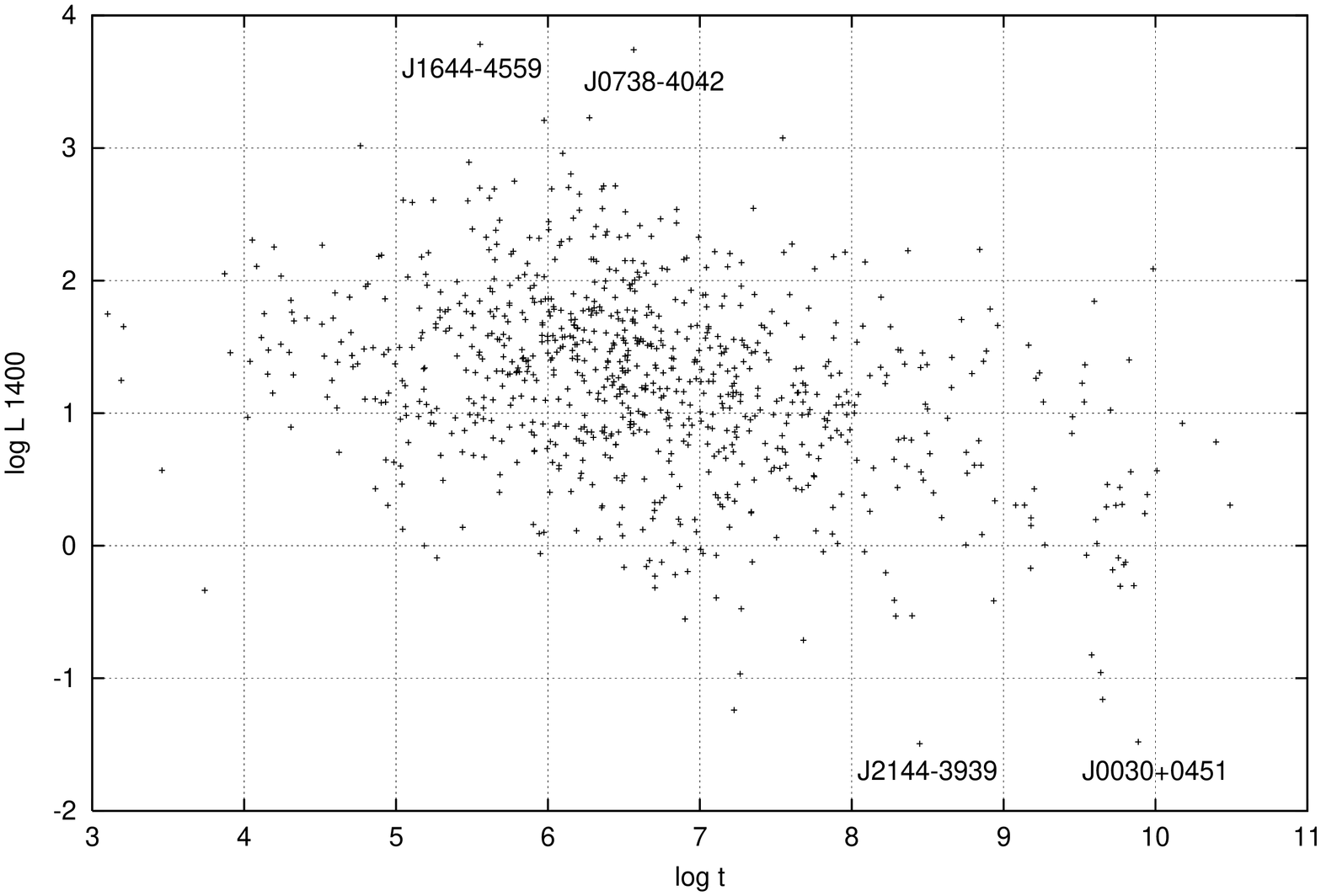,width=16cm,angle=-90}}
{Figure 7a. Luminosities of PSRs at 1400 MHz as a function of  
characteristic age for 853 PSRs
(5 PSRs in MC is included).} 
\end{figure*}

\begin{figure*}
\centerline{\psfig{file=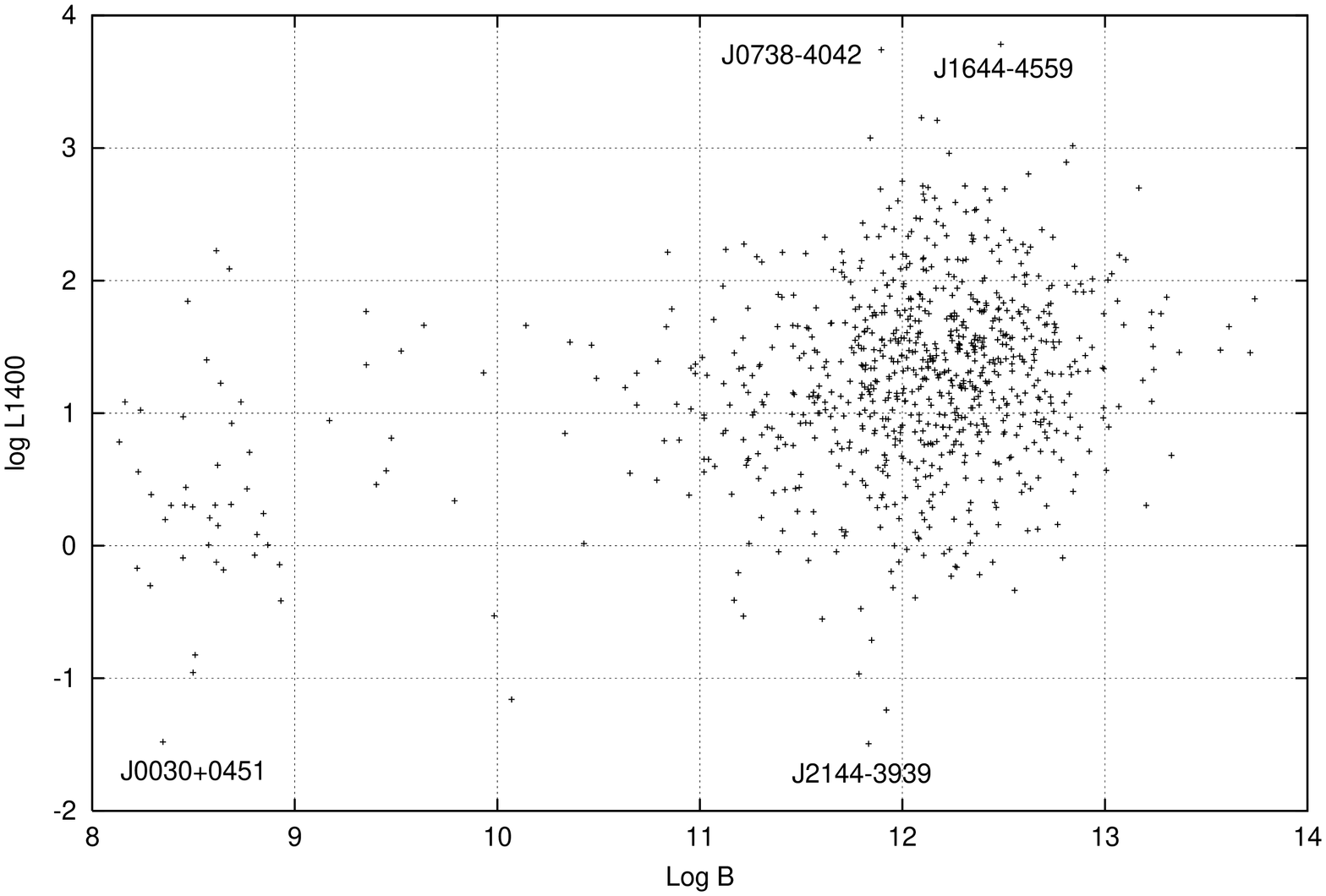,width=16cm,angle=-90}}
{Figure 7b. Luminosity at 1400
MHz vs magnetic field for 853 
PSRs.} 
\end{figure*}

\newpage
\clearpage

\begin{figure*}
\centerline{\psfig{file=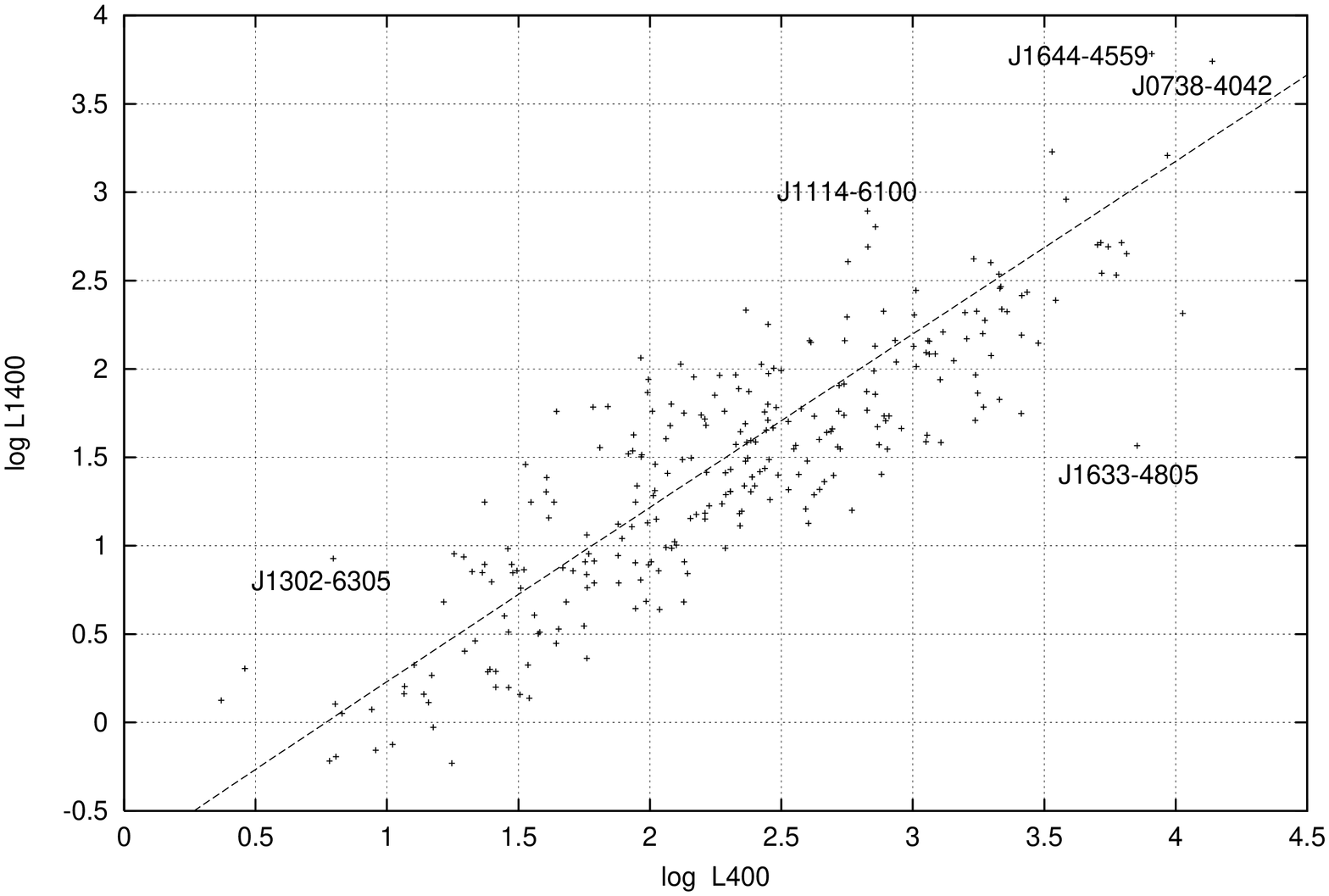,width=16cm,angle=-90}}
{Figure 8a. Log L$_{1400}$-Log L$_{400}$
diagram for 273 PSRs with $\tau$ $<$ 10$^7$ years.}
\end{figure*}

\begin{figure*}
\centerline{\psfig{file=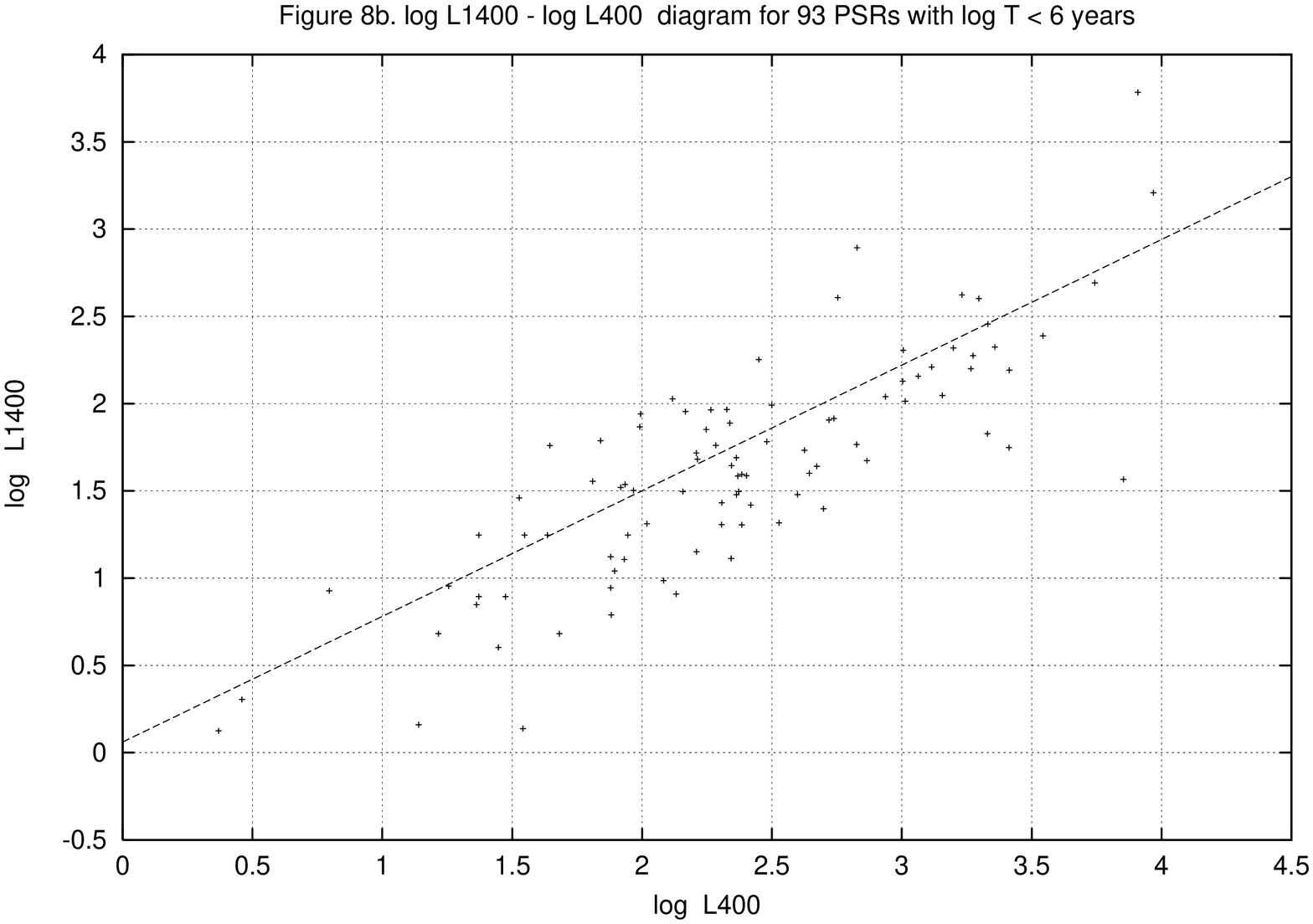,width=16cm,angle=-90}}
{Figure 8b. Log L$_{1400}$-Log L$_{400}$
diagram for 93 PSRs with $\tau$ $<$ 10$^6$ years.}
\end{figure*}

\begin{figure*}
\centerline{\psfig{file=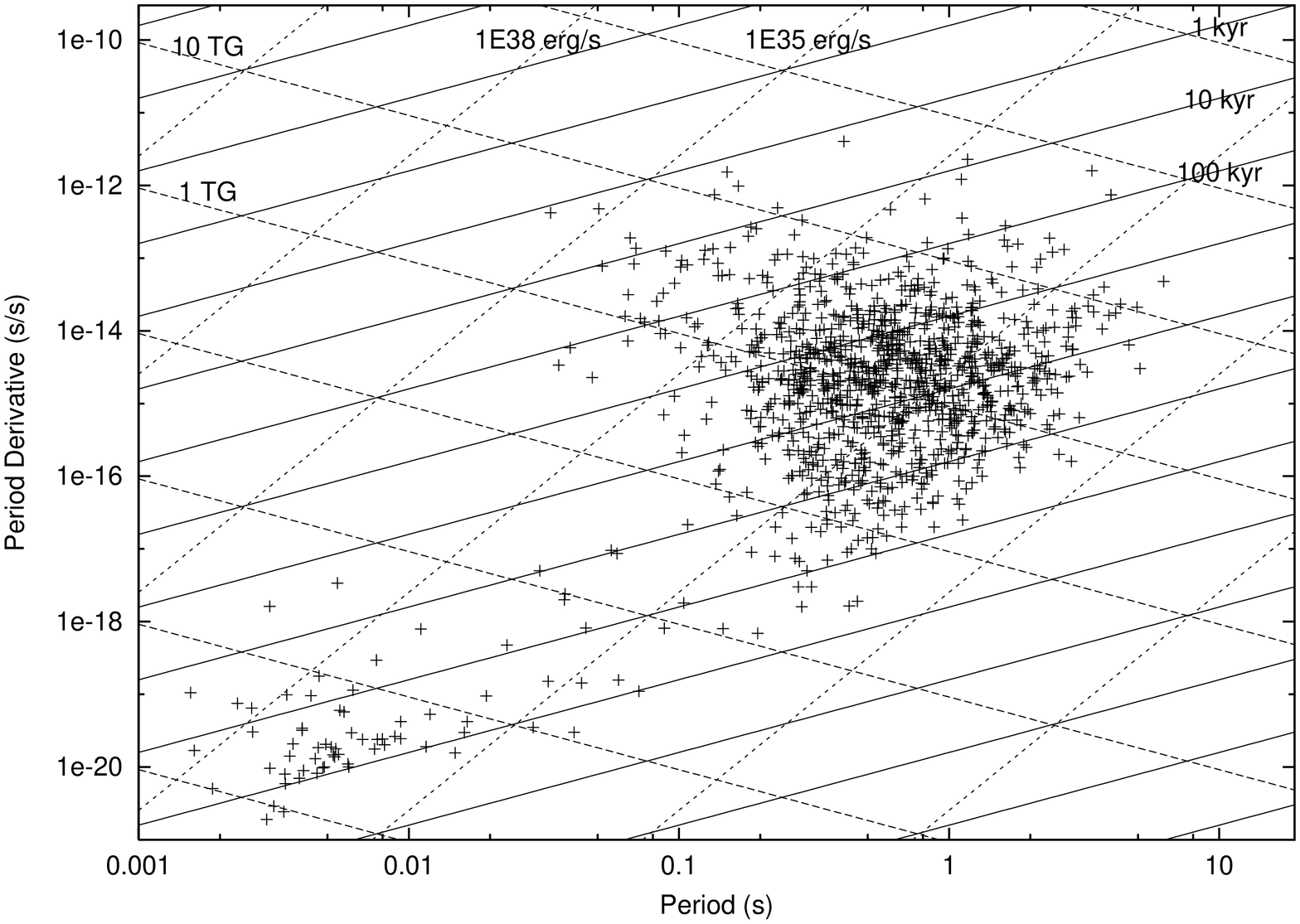,width=16cm,angle=-90}}
{Figure 9. The $P$-$\dot{P}$ Diagram for 1191 PSRs. The constant 
characteristic age lines are given by the formula $\tau$=P/2\.{P}, and 
the lines for 1, 10 and 100 kyr are indicated. The constant magnetic 
dipole field lines are given by the formula B=3.3 10$^{19}$(P\.{P})$^{1/2}$, 
and the lines for 1 and 10 TG are indicated. The constant rate of energy 
loss lines are given by the formula \.{E}=3.94 10$^{46}$\.{P}/P$^3$ and 
the lines for 10$^{38}$ and 10$^{35}$ are indicated.} 
\end{figure*}

\end{document}